\documentclass[conference]{IEEEtran}
\IEEEoverridecommandlockouts
\usepackage{cite}
\usepackage{soul}
\usepackage{amsmath,amssymb,amsfonts}
\usepackage{algorithm,algorithmic}
\usepackage{graphicx}
\usepackage{textcomp}
\usepackage{xcolor}
\def\BibTeX{{\rm B\kern-.05em{\sc i\kern-.025em b}\kern-.08em
    T\kern-.1667em\lower.7ex\hbox{E}\kern-.125emX}}
\begin{document}

\title{Interoperable rApp/xApp Control over O-RAN for Mobility-aware Dynamic Spectrum Allocation
%{\footnotesize \textsuperscript{*}Note: Sub-titles are not captured in Xplore and should not be used}
\thanks{This work was supported in part by UNITY-6G Project, funded by the European Union’s HORIZON-JU-SNS-2024 program under Grant No 101192650; and in part by the 6G-Cloud Project funded from the European Union’s HORIZON-JU-SNS-2023 program under Grant No 101139073.}
}

\author{\IEEEauthorblockN{Anastasios Giannopoulos}
\IEEEauthorblockA{\textit{R\&D Department} \\
\textit{Four Dot Infinity}\\
Athens, Greece \\
angianno@fourdotinfinity.com}
\and
\IEEEauthorblockN{Sotirios Spantideas}
\IEEEauthorblockA{\textit{R\&D Department} \\
\textit{Four Dot Infinity}\\
Athens, Greece \\
sospanti@fourdotinfinity.com}
\and
\IEEEauthorblockN{Maria Lamprini Bartsioka}
\IEEEauthorblockA{\textit{R\&D Department} \\
\textit{Four Dot Infinity}\\
Athens, Greece \\
mbarts@fourdotinfinity.com}
\and
\IEEEauthorblockN{Panagiotis Trakadas}
\IEEEauthorblockA{\textit{R\&D Department} \\
\textit{Four Dot Infinity}\\
Athens, Greece \\
ptrak@fourdotinfinity.com}
}

\maketitle

\begin{abstract}
Open Radio Access Networks (O-RAN) enable the disaggregation of radio access functions and the deployment of control applications across different timescales. However, designing interoperable control schemes that jointly exploit long-term traffic awareness and near-real-time radio resource optimization remains a challenging problem, particularly under dense multi-cell interference and heterogeneous service demands. This paper proposes an interoperable rApp/xApp-driven dynamic spectrum allocation (DSA) framework for O-RAN, based on a graph-theoretic formulation of physical resource block (PRB) assignment. The proposed architecture leverages a non-real-time radio intelligent controller (Non-RT RIC) rApp to predict aggregated traffic evolution and generate high-level spectrum policies at the minutes timescale, while a near-real-time RIC (Near-RT RIC) xApp constructs a user-centric conflict graph and performs fairness-aware PRB allocation at sub-second timescales. To mitigate persistent user starvation, a conflict-aware modified proportional fair (MPF) scheduling mechanism is applied, enabling controlled interference-free PRB time-sharing. Extensive simulation results demonstrate that the proposed framework significantly improves the PRB assignment success rate (above 90\%) and service-share fairness (above 85\%) across different channel configurations and user demands, while maintaining architectural separation and rApp/xApp interoperability in accordance with O-RAN principles.
\end{abstract}


\section{Introduction}

%0.75 page

%Goal: Motivate the problem, define the gap, state contributions crisply.

%Content:

\IEEEPARstart{O}{pen} Radio Access Networks (O-RAN) have gained significant attention in recent years due to their flexibility, enabled by open interfaces and the softwarization and virtualization of RAN components \cite{alliance2020}. Towards sixth-generation (6G) systems, monolithic base stations (BSs) are replaced by disaggregated and programmable components, namely the O-Radio Unit (O-RU) for radio signal processing, the O-Distributed Unit (O-DU) for real-time lower-layer functions, and the O-Central Unit (O-CU) for non-real-time upper-layer operations. O-RAN behavior is dynamically controlled by the Near-Real-Time RAN Intelligent Controller (Near-RT RIC), which hosts xApps operating at sub-second time scales and interacting with O-CU/O-DU nodes through the standardized E2 interface to optimize functions such as traffic steering, load balancing, and energy-efficient radio control \cite{giannopoulos2025comix, hoffmann2023open}. Complementarily, the Non-Real-Time RIC (Non-RT RIC) operates at longer time scales and supports rApps responsible for AI/ML model training, policy management, and quality-of-service assurance \cite{bonati2021intelligence,giannopoulos2025ai}.

Although the O-RAN architecture provides concrete advances toward 6G networks, mobile operators remain cautious in its adoption due to the increased system complexity introduced by RAN disaggregation and the need for continuous parameter configuration. In this context, AI/ML-based control of key RAN parameters has gained significant interest, particularly in dense deployments where spectrum resources are scarce. Heterogeneous O-RAN scenarios with coexisting macro and micro O-RUs operating over the same spectrum exacerbate inter-cell interference, making dynamic frequency reuse and scheduling essential. These functions are executed at the O-DU, and prior works have investigated ML-driven dynamic spectrum allocation (DSA) and scheduling schemes based on proportional fairness (PF), water-filling (WF), and round-robin (RR) principles \cite{polese2022colo}. In O-RAN, such ML models are typically deployed as xApps in the Near-RT RIC, with scheduling policies enforced at the O-DU via the E2 interface \cite{alliance2020ai}.

The decisions implemented by existing ML-based xApps typically rely on instantaneous radio conditions and do not exploit higher-level contextual information, such as user mobility trends or temporal traffic patterns, which can be extracted at the Non-RT RIC. As a result, interoperability between rApps and xApps becomes critical, since the enriched, long-term network view available at the management plane can enable more informed RAN configurations and improved spectrum allocation decisions. While interoperable rApp/xApp frameworks have been explored for specific use cases, such as joint traffic steering and energy saving in multi-vendor environments \cite{akman2024energy}, or Coverage and Capacity Optimization (CCO) from a global management perspective \cite{10651558}, the use of long-term traffic awareness at the Non-RT RIC to systematically guide near-real-time spectrum allocation in the Near-RT RIC remains largely unexplored. This paper addresses this gap by:

\begin{itemize}
    \item Extensively presenting a generalizable end-to-end interoperable rApp/xApp control framework for mobility-aware DSA in O-RAN, considering the relevant time-scales and the information flow between the architectural O-RAN components. Analytic workflows are also provided to showcase rApp/xApp closed-loop interoperability.
    \item Developing a graph-based interference modeling to enable a conflict-constrained spectrum allocation coloring mechanism (xApp-level), which periodically receives predictive traffic-driven optimal configuration parameters from the rApp level.
    \item Implementing a policy-guided modified proportional fair (MPF) mechanism as a post-coloring step to enable time-sharing of spectral resources and enhance fairness among the users.
    \item Evaluating the proposed interoperable rApp/xApp framework in  various scenarios that consider the operation of macro/micro RUs and multiple mobile users under severe interference conditions. Evaluation considers both training and inference results and the algorithm performance is assessed in terms of service success and fairness scores. 
\end{itemize}

%Spectrum scarcity and inter-cell interference in dense multi-cell O-RAN deployments (macro + micro RUs, mobility).

%Limitations of:

%Static or purely DU-local scheduling,

%Monolithic ML approaches without architectural separation.

%Key gap: Lack of interoperable rApp/xApp frameworks where:

%Long-term traffic awareness informs Near-RT graph-based spectrum decisions.

%End with explicit contributions (bullet list):

%An interoperable rApp–xApp control framework for dynamic spectrum allocation in O-RAN.

%A graph-based interference modeling and coloring mechanism executed at the xApp.

%A policy-guided modified proportional fair (PF) scheduler enforced at the DU.

%A multi-cell evaluation with macro/micro RUs and mobile users.

\section{System Model and O-RAN Architecture}

\begin{figure}[t]
\centerline{\includegraphics[trim={2.1cm 1.5cm 2.5cm 1.5cm},clip,width=\columnwidth]{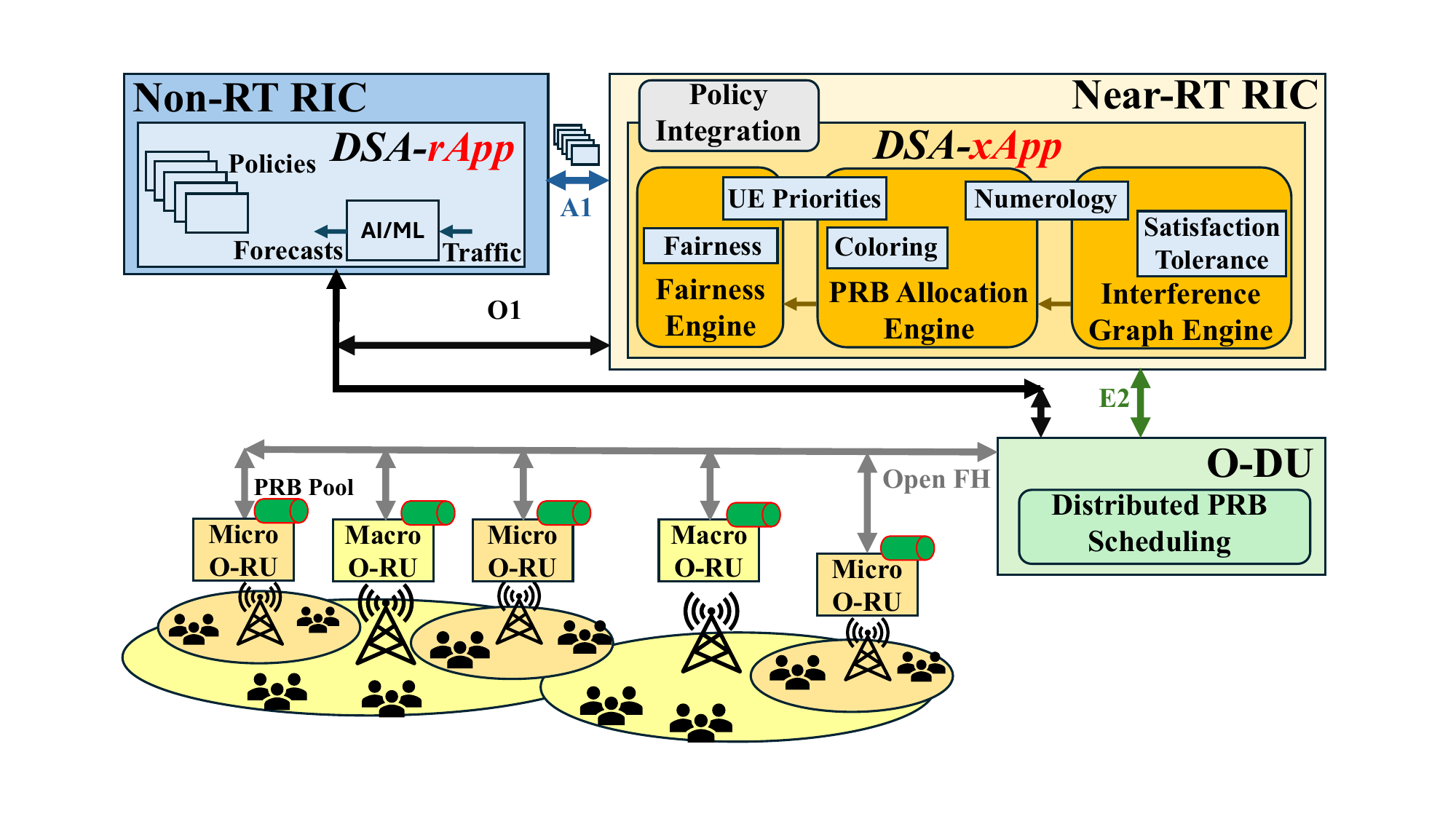}}
\caption{Interoperable rApp/xApp functionalities and heterogeneous O-RAN architecture with multiple co-operating O-RUs.}
\label{fig:fig1}
\end{figure}

We consider a multi-cell O-RAN-compliant system composed of a Non-RT RIC, a Near-RT RIC, an O-DU, and a set of heterogeneous macro-area and micro-area O-RUs, as illustrated in Fig.~\ref{fig:fig1}. The system operates over a shared spectrum pool and serves a set of mobile user equipments (UEs).

\subsection{Network Entities and Radio Resources Model}

Let $\mathcal{R} = \{1,2,\dots,R\}$ denote the set of O-RUs, where each RU $r \in \mathcal{R}$ corresponds either to a macro-cell or to a micro-cell. Let $\mathcal{U} = \{1,2,\dots,U\}$ denote the set of active UEs served by the network. The available radio spectrum is partitioned into a finite set of physical resource blocks (PRBs) drawn from the set $\mathcal{P}=\{1,2,\dots,P\}$. Each RU $r$ can transmit at a maximum sum-power level $P_r^{\text{max}}$, as reflected below:

\begin{equation}
    \sum_{p \in \mathcal{P}} P_{r,p}\leq P_r^{\text{max}}
\end{equation}

\noindent where $P_{r,p}$ is the power level of PRB $p$ by RU $r$. Also, each PRB-specific power level must exceed a minimum value (for beacon transmissions), i.e., $P_{r,p} \geq P^{\text{min}}$.

\subsubsection{Radio Spectrum Slicing Model}

The physical-layer resource grid is defined according to the 5G New Radio (NR) specifications \cite{lien20175g}. Let $B$ denote the total system bandwidth and $B_G$ the two-sided guard-band. The effective usable bandwidth is therefore $B_{\text{eff}} = B - 2B_G$. The sub-carrier spacing is determined by the selected NR numerology $\mu$ as $\Delta f = 15 \cdot 2^{\mu}$ (in kHz). Each PRB consists of 12 sub-carriers, and hence the PRB bandwidth is given by $W = 12 \cdot \Delta f$. The total number of available PRBs is then computed as \cite{lien20175g}:

\begin{equation}
P = \left\lfloor \frac{B - 2B_G}{180 \cdot 2^{\mu}} \right\rfloor
\end{equation}

\noindent where $\lfloor \cdot \rfloor$ denotes the floor operator. The set $\mathcal{P}$ is dynamically reused across cells subject to inter-cell interference constraints.

\subsubsection{User Mobility Model}

UEs are mobile and their positions evolve over time according to a two-dimensional random-walk mobility model \cite{chiang20042}. The position of UE $u$ at time $t$ is denoted by $\mathbf{l}_u(t) = \big[x_u(t), y_u(t)\big]^{\top}$. At each mobility update interval $\Delta t$, the position of UE $u$ is updated as:

\begin{equation}\label{eq:mobility}
\mathbf{l}_u(t+\Delta t) =
\mathbf{l}_u(t) +
v_u \Delta t
\begin{bmatrix}
\cos\theta_u(t) \\
\sin\theta_u(t)
\end{bmatrix}
\end{equation}

\noindent where $v_u$ denotes the UE speed and $\theta_u(t)$ is a direction angle selected at each update. To avoid unrealistic oscillatory motion, UE direction updates are temporally correlated \cite{chiang20042}. Specifically, the direction angle evolves smoothly over time according to a correlated random walk, ensuring bounded heading changes between successive mobility updates. This prevents abrupt direction reversals and results in realistic UE trajectories.

\subsubsection{Channel and Interference Model}

The dynamic location of UE $u$ and the propagation effects directly affect channel gains, received power, and inter-cell interference. Let $D_{r,u}(t)$ denote the distance between RU $r$ and UE $u$. The large-scale path loss between RU $r$ and UE $u$ is modeled as:

\begin{equation}
\mathrm{PL}_{r,u}(t) = K_r \, D_{r,u}(t)^{-\alpha_r}
\end{equation}

\noindent where $\alpha_r$ is the path-loss exponent and $K_r$ is a cell-type-dependent constant, allowing different propagation characteristics for macro and micro cells. The overall channel gain on PRB $p$ is given by:

\begin{equation}
g_{r,u,p}(t) = \mathrm{PL}_{r,u}(t) \cdot h_{r,u,p}(t)
\end{equation}

\noindent where $h_{r,u,p}(t)$ models small-scale fading and is assumed to be Rayleigh distributed \cite{sklar2002rayleigh}. We define the binary PRB assignment variable $x_{u,p}^{(r)}(t)=1$ when PRB $p$ is assigned to UE $u$ by RU $r$ at time $t$, and $0$ otherwise. Each UE is served by a single RU, and each PRB can be allocated to at most one UE per RU and scheduling interval. The signal-to-interference-plus-noise ratio (SINR) experienced by UE $u$ on PRB $p$ from RU $r$ is given by:

\begin{equation}
\mathrm{SINR}_{u,p}^{(r)}(t)=
\frac{x_{u,p}^{(r)}(t) \cdot P_{r,p} \cdot g_{r,u,p}(t)}
{\sum\limits_{\substack{k \in \mathcal{R} \\ k \neq r}} P_{k,p}\, g_{k,u,p}(t) + \sigma^2}
\end{equation}

\noindent where $\sigma^2$ denotes the thermal noise power.

\subsubsection{User Satisfaction Model}

The total instantaneous achievable data rate of UE $u$ is computed using the Shannon capacity formula:

\begin{equation}
R_u(t) = \sum_{r \in \mathcal{R}} \sum_{p \in \mathcal{P}} x_{u,p}^{(r)}(t) \cdot W \log_2 \big(1 + \mathrm{SINR}_{u,p}^{(r)}(t)\big)
\end{equation}

Each UE $u$ has a target data rate demand $d_u$. A UE is considered satisfied if its achieved rate meets a minimum fraction of its demand, defined by a satisfaction tolerance margin $\eta_u \in [0,1]$, as $R_u(t) \geq (1-\eta_u) d_u$. This satisfaction condition is used by the xApp to regulate interference tolerance, PRB allocation, and fairness decisions.

\subsection{O-RAN Components and their Functional Roles}\label{sec:roles}

The proposed system follows a strict functional and temporal separation across O-RAN entities, enabling interoperable and scalable, mobility-aware and dynamic spectrum allocation. The key components are described as follows:

\subsubsection{Non-RT RIC and rApp Operation}

A dynamic spectrum allocation rApp (DSA-rApp) is deployed at the Non-RT RIC and operates at a minutes-level timescale \cite{levis2025sleepy}. Based on historical traffic measurements collected via the O1 interface, the rApp performs traffic prediction for each RU over a future horizon $\Delta T$ (e.g., anticipated traffic within the next 15-minutes), based on a historical lookback window of length $L$ \cite{spantideas2024smart}. Specifically, the predicted traffic load of RU $r$ is given by:

\begin{equation}
\hat{\lambda}_r(t+\Delta T)
=
\mathcal{F}_{\mathrm{ML}}\big(\lambda_r(t),\lambda_r(t-1),\dots,\lambda_r(t-L) \big)
\end{equation}

\noindent where $\mathcal{F}_{\mathrm{ML}}(\cdot)$ denotes a generic AI/ML-based forecasting function. Using these predictions, the rApp generates high-level policy descriptors that configure the behavior of the Near-RT RIC's xApp modules. These policies are combined in a policy profile (e.g. as JSON file) which is conveyed to the Near-RT RIC through the A1 interface. This policy profile considered for the proposed DSA optimization scenario includes:

\begin{itemize}
    \item \textbf{Policy 1:} This policy includes UE priority classes (e.g., 'High' or 'Low') defined by Service-Level Agreement (SLA), indicating that some of the users must be prioritized in the PRB assignment process (e.g., gold vs standard users),
    \item \textbf{Policy 2:} This policy (selected by the Non-RT RIC) defines the algorithmic approach to be used for final PRB allocation to ensure equitable spectrum access across UEs. These algorithms include Round Robin (RR), conventional Proportional Fair (PF) \cite{kim2005proportional}, or a modified PF (MPF) scheme proposed in this study to guarantee time-sharing PRB allocation.
    \item \textbf{Policy 3:} It defines the SLA-provided interference tolerance margins per UE (i.e., $\eta_u$) used during the interference graph construction. This UE-specific satisfaction tolerance indicates the ratio to which a certain user can accept demand degradation, hence affecting the density (number of conflicts) of the interference graph.
    \item \textbf{Policy 4:} Another rApp-driven configuration is selected by the Non-RT RIC and indicates the DSA algorithm (simplistic, heuristic, or learning-based) to be used for the interference graph coloring such as Greedy Coloring \cite{sipayung2022implementation}, Welsh-Powell Coloring \cite{welsh1967upper}, or Degree of Saturation (DSatur) Coloring \cite{san2012new}.  
    \item \textbf{Policy 5:} This policy is rApp prediction-driven and defines the spectrum aggressiveness level (or numerology), which in turn affects the number of available PRBs \cite{lien20175g}. Specifically, according to this policy, the total channel bandwidth is segmented dynamically (in the minutes-to-hours scale) into PRBs based on the highest (across RUs) predicted traffic. This is achieved by setting the numerology $\mu$ based on whether the expected traffic intensity is low (e.g., after mid-night hours), moderate or high (e.g., during the morning or afternoon). The expected worst-case traffic $\hat{\lambda}_{worst}(t+\Delta T)$ for the next $\Delta T$ minutes is computed as:

    \begin{equation}
        \hat{\lambda}_{worst}(t+\Delta T) = \max_{r \in \mathcal{R}} \Bigl( \hat{\lambda}_r(t+\Delta T) \Bigr)
    \end{equation}
\end{itemize}

\subsubsection{Near-RT RIC and xApp Operation}

A DSA-xApp operates within the Near-RT RIC at a sub-second timescale. The xApp is responsible for real-time spectrum coordination based on instantaneous radio conditions and rApp-provided policies. The DSA-xApp supports a three-module optimization scheme, including an Interference Graph Engine, a PRB Allocation Engine, and a Fairness Engine. Firstly, in the \textit{Interference Graph Engine} \cite{zhang2013interference}, PRB assignment conflicts between user pairs (due to interference) are modeled through a dynamic interference graph $\mathcal{G}(t)=\big(\mathcal{V}(t),\mathcal{E}(t)\big)$,
where each vertex $u \in \mathcal{V}(t)$ represents an active UE $u \in \mathcal{U}$.
An undirected edge $(u,v) \in \mathcal{E}(t)$ indicates that UEs $u$ and $v$ cannot occupy the same PRB at time $t$. A conflict edge exists (i) between all same-RU users (because they cannot occupy a common PRB), and (ii) between two different-RU users if assigning a common PRB $p \in \mathcal{P}$ to both users causes at least one of them to violate its satisfaction condition. Formally, an edge $(u,v)$ exists if $\exists p \in \mathcal{P} | R_{u,p}(t) < (1-\eta_u)d_u \text{ or } R_{v,p}(t) < (1-\eta_v)d_v$,
where $R_{u,p}(t)$ and $R_{v,p}(t)$ denote the achievable rates of users $u$ and $v$, respectively. Under this formulation, graph $\mathcal{G}(t)$ is dynamically updated based on users' mobility, interference conditions, and demand levels.

Based on the constructed graph, the DSA-xApp's \textit{PRB Allocation Engine} formulates the frequency resource allocation as a graph coloring problem \cite{ge20222}, where each color corresponds to a PRB and adjacent vertices cannot share the same color. Let $c_u(t) \in \mathcal{P}$ denote the PRB assigned to UE $u$ at time $t$. The coloring must satisfy $c_u(t) \neq c_v(t)$, for all edges $(u,v) \in \mathcal{E}(t)$.

In addition to conflict avoidance, the xApp incorporates a \textit{Fairness Engine}, which applies criteria to prioritize users according to both instantaneous channel quality and long-term fairness \cite{kim2005proportional}. To ensure that uncolored UEs are not constantly unsatisfied, this engine can apply fairness schemes to ensure that PRBs are time-shared equitably among users, while respecting the conflicts induced by the interference graph. The DSA-xApp's outcome is a PRB-to-UE assignment matrix that is forwarded to the O-DU via the E2 interface for enforcement.

\subsubsection{O-DU and Scheduling Enforcement}

The O-DU is responsible for enforcing the PRB assignments received from the Near-RT RIC. It applies the xApp-generated PRB allocation decisions across the connected O-RUs through the open fronthaul interface. The O-DU does not perform spectrum optimization but executes the scheduling decisions computed by the xApp. Overall, this functional split ensures clear and interoperable separation between long-term intelligence (rApp), near-real-time optimization (xApp), and real-time execution (O-DU), in alignment with O-RAN design principles.

\section{Interoperable rApp/xApp-driven DSA Control Framework}

This section presents the proposed interoperable DSA framework based on coordinated rApp/xApp operation within the O-RAN architecture. The framework exploits long-term traffic awareness at the Non-RT RIC and near-real-time graph-based optimization at the Near-RT RIC, forming a closed control loop for fairness-aware PRB allocation in multi-cell environments.

\begin{algorithm}[t]
\caption{Policy-Guided Graph Coloring}
\begin{algorithmic}[1]
\STATE \textbf{Input:} $\mathcal{G}(t)$, $\{w_u\}$, $\mathcal{P}$
\STATE Sort UEs in descending order of weighted degree
\FOR{each UE $u$ in sorted order}
    \FOR{each PRB $p \in \mathcal{P}$}
        \IF{$p$ does not conflict with neighbors of $u$}
            \STATE Assign $c_u \leftarrow p$
            \STATE \textbf{break}
        \ENDIF
    \ENDFOR
\ENDFOR
\STATE \textbf{Output:} Colored graph $\mathcal{G}'(t)$
\end{algorithmic}
\label{alg1}
\end{algorithm}

\subsection{rApp-Driven Traffic Prediction and Policy Configuration} 

\begin{algorithm}[t]
\caption{Modified Proportional Fair for Post-Coloring Scheduling}
\begin{algorithmic}[1]
\STATE \textbf{Input:} $\mathcal{G}'(t)$, $\{w_u\}$, $\mathcal{P}$
\STATE Initialize PRB usage according to $\mathcal{G}'(t)$
\FOR{each uncolored UE $u \in \mathcal{U}_{unc}(t)$}
    \STATE Compute MPF metric $M_u(t)$ over candidate PRBs
    \STATE Compute $p^* = \arg\max_{p\in \mathcal{P}} M_u(t)$
    \IF{$p^*$ is assigned to UE $v$ \textbf{and} $M_u(t)>M_v(t)$
    }
        \STATE UE $u$ is scheduled on PRB $p$
    \ELSE
        \STATE UE $v$ is scheduled on PRB $p$
    \ENDIF
\ENDFOR
\FOR{each UE $u \in \mathcal{U}$}
    \STATE Update $\bar{R}_u(t)$ using EWMA \eqref{EWMA}
\ENDFOR
\STATE \textbf{Output:} Final PRB allocation $\{c_u\}$
\end{algorithmic}
\label{alg2}
\end{algorithm}

The DSA-rApp is responsible for long-term traffic intelligence and policy configuration. Its operation follows the steps below:

\textbf{Step 1: Continuous Traffic Monitoring} The rApp collects historical key performance measurements (KPMs) via the O1 interface, including aggregated traffic load per RU $\lambda_r(t)$ in terms of UE density, mobility statistics, and long-term throughput and satisfaction indicators.
These measurements are stored in a time-series database (or data lake), which is updated periodically.

\textbf{Step 2: Offline rApp Training:} Using historical traffic traces, the rApp trains a traffic prediction model offline (e.g., LSTM network). Offline training is performed infrequently and does not affect online RAN operation. Offline training is performed infrequently and does not affect online RAN operation.

\textbf{Step 3: Minutes-level rApp Inference:} At runtime, the rApp performs inference at a minutes-level timescale to estimate upcoming traffic conditions for each RU. These predictions are exploited to dynamically configure the channel segmentation policy into PRBs (Policy 5) and/or select the Coloring (Policy 4) and Fairness (Policy 2) schemes to be used by the xApp, based on historical performance of different schemes for previous traffic density instances.

\textbf{Step 4: Policy Construction and Dissemination:} Based on the predicted traffic, the rApp constructs high-level policy descriptors that configure the behavior of the xApp. Each policy is represented as a tuple $\boldsymbol{\pi} =
\big(
\boldsymbol{\pi}_1,\,
\boldsymbol{\pi}_2,\,
\boldsymbol{\pi}_3,\,
\boldsymbol{\pi}_4,\,
\boldsymbol{\pi}_5
\big)$, where $\boldsymbol{\pi}_1$ is an SLA-based UE priorities vector $\boldsymbol{w}=\{w_u\}$, $\boldsymbol{\pi}_2$ specifies the selected fairness scheme, $\boldsymbol{\pi}_3$ is an SLA-based UE satisfaction tolerance vector $\boldsymbol{\eta}=\{\eta_u\}$, $\boldsymbol{\pi}_4$ determines the selected coloring scheme, and $\boldsymbol{\pi}_5$ defines the selected numerology based on traffic predictions.

The policy descriptors are transmitted to the Near-RT RIC via the A1 interface. Hence, the rApp does not issue direct PRB commands, but only configures the decision space of the xApp's internal engines.

\subsection{xApp-Based Fairness-aware PRB Assignment Algorithm}

\begin{figure}[t]
\centerline{\includegraphics[trim={1.4cm 0.2cm 1.3cm 0.5cm},clip,width=\columnwidth]{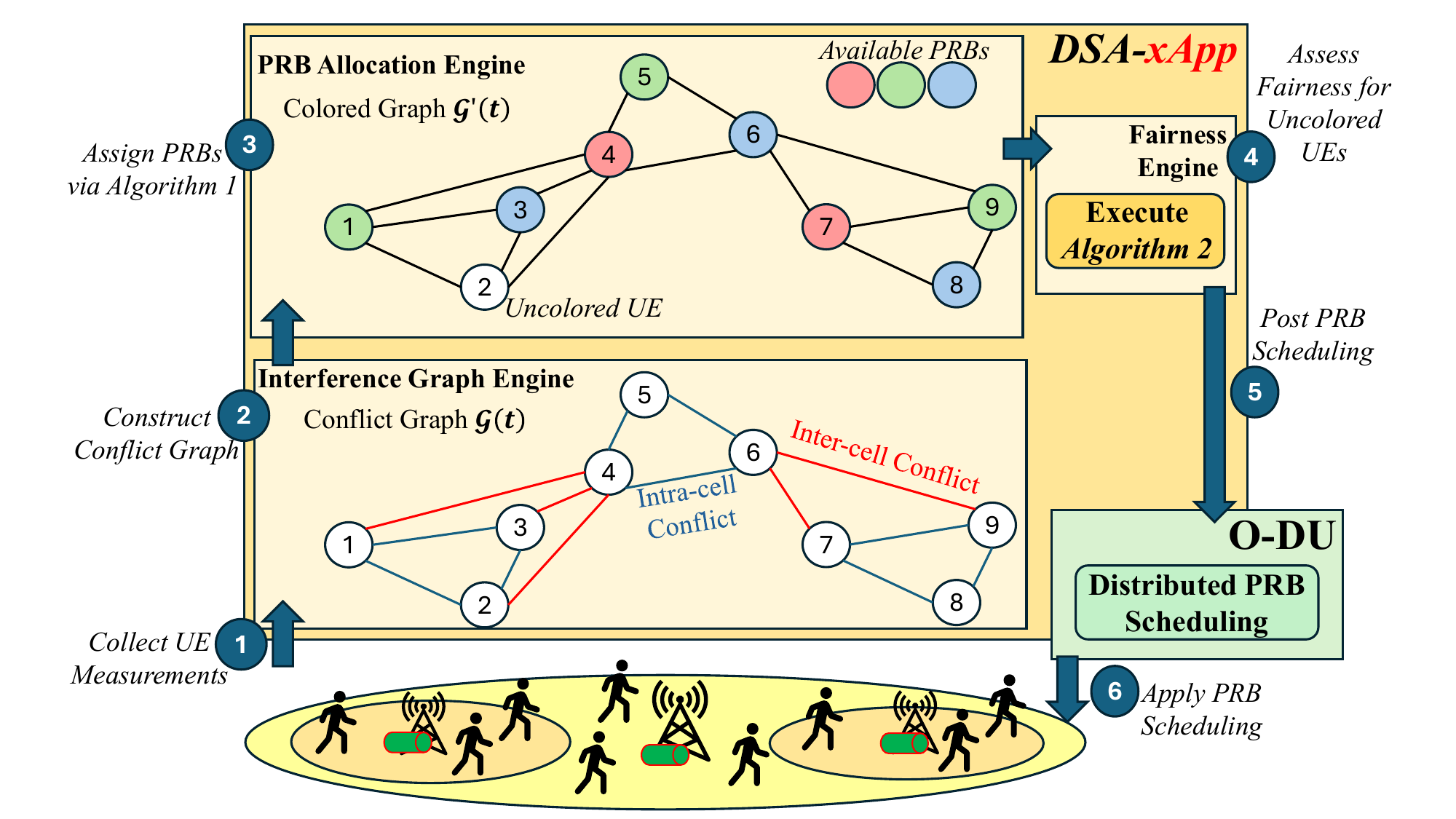}}
\caption{Graph-based continuous optimization cycle of DSA-xApp.}
\label{fig:fig2}
\end{figure}

The DSA-xApp executes dynamic PRB allocation under rApp-provided policies. The closed-loop xApp control is visualized in Fig.~\ref{fig:fig2}. The following phases are implemented in the Near-RT RIC:

\textbf{Phase 1: Policy Integration:} Upon receiving policy $\boldsymbol{\pi}$, the Policy Integration module configures DSA-xApp's three internal engines. Policy parameters directly affect conflict thresholds, fairness weights, and coloring strategy.

\textbf{Phase 2: Interference Graph Construction} Using $\boldsymbol{\pi}_3$ and $\boldsymbol{\pi}_5$, the xApp first constructs a user-centric conflict graph $\mathcal{G}(t)$ to represent the time-specific PRB allocation constraints across all O-RAN users.

\textbf{Phase 3: PRB Allocation Optimization:} Based on $\boldsymbol{\pi}_1$, $\boldsymbol{\pi}_4$ and $\boldsymbol{\pi}_5$, the PRB allocation problem is formulated as a weighted graph coloring problem:

\begin{align}
\max_{\{c_u\}} \quad
& \sum_{u\in\mathcal{U}} w_u \cdot \min(R_{u,c_u}(t),d_u) \\
\text{s.t.} \quad
& c_u \neq c_v, \quad \forall (u,v)\in\mathcal{E}(t), \\
& c_u \in \mathcal{P}, \quad \forall u\in\mathcal{U}, \\
& \sum_{p \in \mathcal{P}} P_{r,p}\leq P_r^{\text{max}}, \quad \forall r\in\mathcal{R} \\
& P_{r,p}\geq P^{\text{min}}, \quad \forall r\in\mathcal{R}, p\in\mathcal{P}
\end{align}

\noindent where the objective function reflects the total weighted and demand-bounded data rate maximization target. Due to its NP-hard nature, the problem is solved using a heuristic coloring algorithm, which is implemented by the PRB Allocation Engine. The generic xApp logic is abstracted in Algorithm~\ref{alg1}.

\textbf{Phase 4: Modified Proportional Fair Scheduling:} Due to conflict constraints imposed by the graph coloring process, a subset of UEs may remain temporarily uncolored in $\mathcal{G}'(t)$. To avoid persistent starvation and ensure long-term fairness, the DSA-xApp applies a conflict-aware modified PF scheduling mechanism as a post-coloring step. The MPF mechanism allows uncolored UEs to time-share PRBs with already-colored UEs. In contrast to classical PF \cite{kim2005proportional}, which operates over all UEs symmetrically, the proposed MPF explicitly accounts for coloring decisions and conflict relationships. For each UE $u$, the MPF metric is defined as:

\begin{equation}
M_u(t) = \frac{R_{u,p}(t)}{\bar{R}_u(t)} \cdot w_u.
\end{equation}

\noindent where $p$ denotes the candidate PRB (either the assigned color for colored UEs or a sharable PRB for uncolored UEs), and $\bar{R}_u(t)$ is the exponentially weighted moving average (EWMA) throughput, updated as:

\begin{equation}\label{EWMA}
\bar{R}_u(t) = (1-\alpha)\bar{R}_u(t-1)
+ \alpha\, R_u(t)
\end{equation}

\noindent where $\alpha \in (0,1]$ is the EWMA smoothing factor and $R_u(t)$ denotes the instantaneous achieved data rate of UE $u$ at time $t$. A smaller value of $\alpha$ emphasizes long-term fairness, while a larger value increases responsiveness to short-term channel variations. Let $\mathcal{U}_{col}(t)$ and $\mathcal{U}_{unc}(t)$ denote the sets of colored and uncolored UEs after the graph coloring phase, respectively. The Fairness Engine executes the MPF as described in Algorithm~\ref{alg2}.

This post-coloring MPF step ensures that uncolored UEs are progressively served over time without destabilizing the coloring solution, thereby balancing fairness, stability, and interference mitigation. The final PRB-to-UE assignment is forwarded to the O-DU via the E2 interface.

\subsection{Proposed Closed-Loop Workflow}

\begin{figure}[t]
\centerline{\includegraphics[trim={0cm 2cm 0cm 0.5cm},clip,width=\columnwidth]{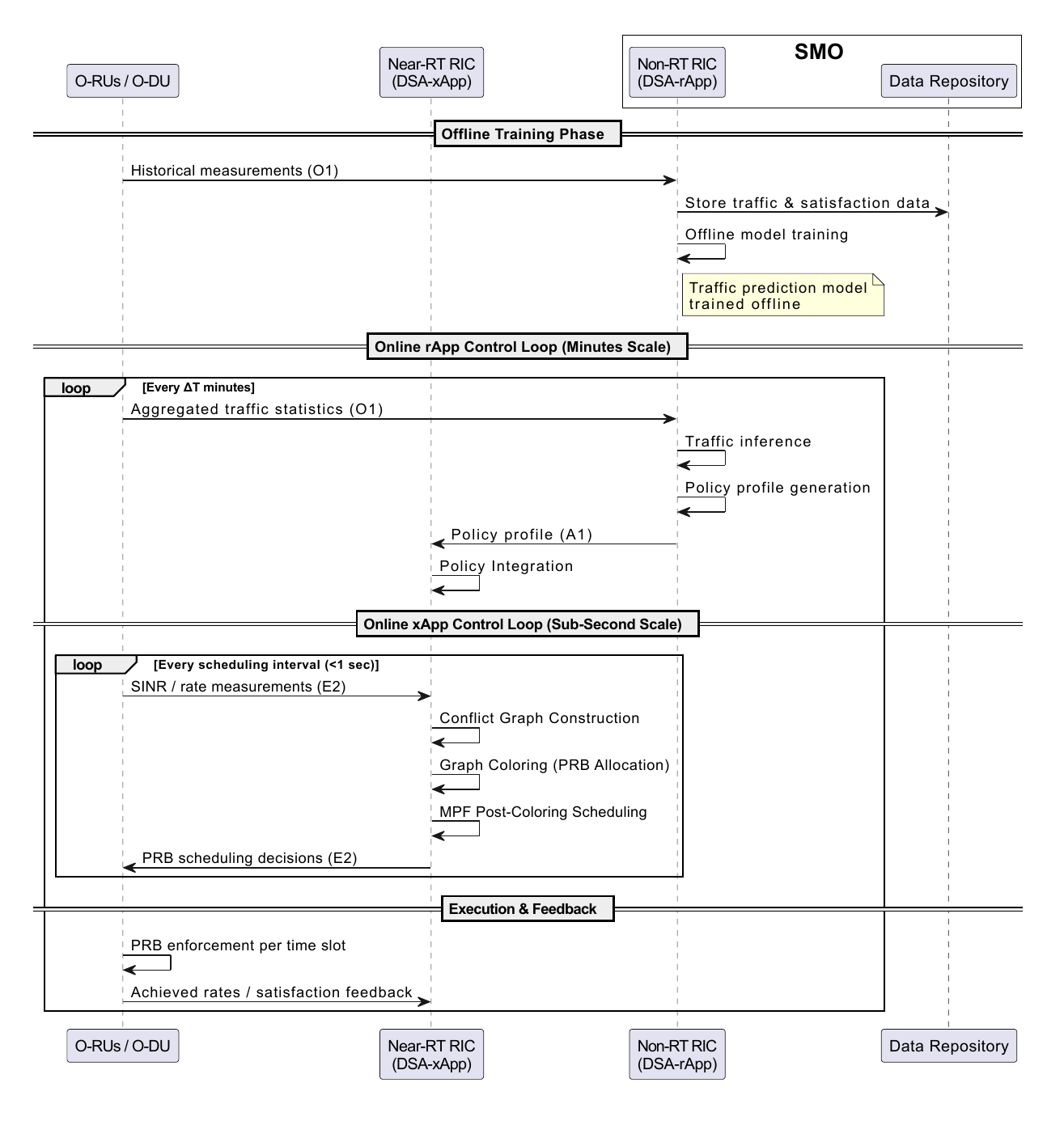}}
\caption{Sequence diagram for closed-loop rApp/xApp-driven DSA workflow.}
\label{fig:uml}
\end{figure}

The proposed interoperable DSA framework operates as a multi-timescale closed-loop control system, integrating long-term intelligence from an rApp with near-real-time optimization from the xApp. Fig.~\ref{fig:uml} illustrates the corresponding sequence diagram, highlighting both the offline and online control phases.

The workflow starts with an offline training phase executed by the DSA-rApp at the Non-RT RIC. During this phase, the rApp collects historical measurements via the O1 interface, which are stored and used to train the traffic prediction model offline. This phase is fully decoupled from real-time RAN operation, ensuring zero impact on network responsiveness.

During network operation, the rApp executes an online control loop at a minutes-level timescale. Based on newly collected measurements, the rApp performs inference using the trained model to predict future traffic conditions. Together with the SLA policies, the prediction results are then combined to produce the xApp policy profile, as described in Section~\ref{sec:roles}. This is disseminated to the Near-RT RIC through the A1 interface and is used to configure the behavior of the xApp's engines without imposing direct scheduling decisions.

By using the rApp policy parameters for configuration, the DSA-xApp operates at a sub-second timescale and forms the inner control loop of the proposed framework. At each scheduling interval, the xApp collects instantaneous SINR and achievable data rate measurements for active UEs via the E2 interface. Based on these inputs, the xApp executes a three-stage inference process, including (i) the satisfaction-aware conflict graph construction, (ii) the coloring-based PRB assignment, and (iii) the post-coloring conflict-constrained MPF mechanism that enables time-sharing for uncolored UEs. The resulting PRB-to-UE scheduling decisions are then forwarded to the O-DU via the E2 interface. Finally, the O-DU enforces the received PRB assignments on a per-time-slot basis and applies them across the connected O-RUs. Execution feedback, including achieved data rates and satisfaction indicators, is continuously reported back to the RIC layers, closing the control loop. This hierarchically nested rApp/xApp loop ensures responsiveness to short-term channel variations while leveraging long-term traffic awareness of the O-RAN environment.

\section{Performance Evaluation}

This section presents numerical outcomes of the DSA-rApp/xApp framework. We consider a controlled multi-cell setup with heterogeneous cells, 5G NR-compliant spectrum, and mobile UEs with diverse service demands.

\subsection{Simulation Setup}

\subsubsection{Topology and System Setup} We consider a dual-type network consisting of one macro O-RU and two micro O-RUs. The macro-cell is located at the origin, while the micro-cells are placed symmetrically at fixed locations (right and left relative to the macro-RU location). Macro-cells operate at higher transmit power than micro-cells, reflecting their larger coverage area. The spectral and network default configuration parameters are tabulated in Table~\ref{tab:sim_params}. The network operates over multiple rApp episodes ($15$-min intervals), with each rApp episode consisting of multiple xApp episodes or scheduling intervals ($1$-sec intervals). From one xApp slot to the next, UE positions are updated based on the mobility model defined in \eqref{eq:mobility}. Equal priority $w_u=1$ (i.e., $\boldsymbol{\pi}_1$) and satisfaction tolerance margin $\eta_u=0$ (i.e., $\boldsymbol{\pi}_3$, indicating that all UEs intent full satisfaction) among all UEs is considered. 

\begin{table}[t]
\caption{Default Network Setup}
\label{tab:sim_params}
\centering
\begin{tabular}{l c}
\hline
\textbf{Parameter} & \textbf{Value} \\
\hline
Number of O-RUs $R$ & $3$ (1 macro, 2 micro) \\
Number of rApp episodes $N_r$ & $96$ \\
Number of xApp episodes $N_x$ & $900$ \\
Total system bandwidth $B$ & $10$ MHz \\
Guard band (one-sided) $B_G$ & $0.25$ MHz \\
NR numerology index $\mu$ & $0-4$ (rApp-defined) \\
Number of active UEs $U$ & $10-160$ (Dataset-defined) \\
Thermal noise power $\sigma^2$ & $-174$ dBm/Hz \\
Macro-cell PRB transmit power $P_{r,p}^{M}$ & $0.1$ W \\
Micro-cell PRB transmit power $P_{r,p}^m$ & $0.01$ W \\
Minimum PRB power $P^{\min}$ & $1$ mW \\
Macro-cell radius & $300$ m \\
Micro-cell radius & $50$ m \\
Macro-cell position & $(0,0)$ \\
Micro-cell positions & $(200,0)$, $(-200,0)$ \\
Path-loss constant $K_r$ & 1 \\
Path-loss exponent (macro) $\alpha_{\mathrm{M}}$ & $2.7$ \\
Path-loss exponent (micro) $\alpha_{\mathrm{m}}$ & $2.8$ \\
UE service types & Mobile broadband \\
UE demand $d_u$ & $\{0.5,1,1.5\}$ Mbps \\
UE speed $v_u$ & $1.5$ m/s \\
UE priority $w_u$ & $1$ \\
UE satisfaction tolerance $\eta_u$ & $0$ \\
xApp Coloring scheme & Welsh-Powell (Algorithm~\ref{alg1}) \\
xApp Fairness scheme & MPF (Algorithm~\ref{alg2}) \\
MPF EWMF parameter $\alpha$ & $0.1$ \\
\hline
\end{tabular}
\end{table}

\subsubsection{Mobile Traffic Dataset}

To model realistic traffic dynamics and train the rApp traffic prediction module, we employ a public 5G traffic dataset \cite{sevgican2020intelligent}. The dataset provides time-series measurements of cellular traffic generated according to 3GPP-compliant assumptions and is sampled at a granularity of one sample every 15 minutes. The time-varying traffic load $\lambda_r(t)$ within each RU represents the number of UEs, and is computed as the aggregated data rate (summed across different UE subscription categories and device types of the dataset) divided by the UE demands \cite{sevgican2020intelligent}. The objective of the rApp ML models is to predict cell-level traffic trends rather than per-UE behavior. The 6-month dataset is split chronologically into training and testing subsets: the first 5 months plus 29 days are used for training rApp LSTM-based traffic predictor, while the 1 day is reserved for testing.

\subsection{rApp/xApp Performance}

\begin{figure}[t]
\centerline{\includegraphics[trim={0.2cm 1.4cm 0cm 0.2cm},clip,width=\columnwidth]{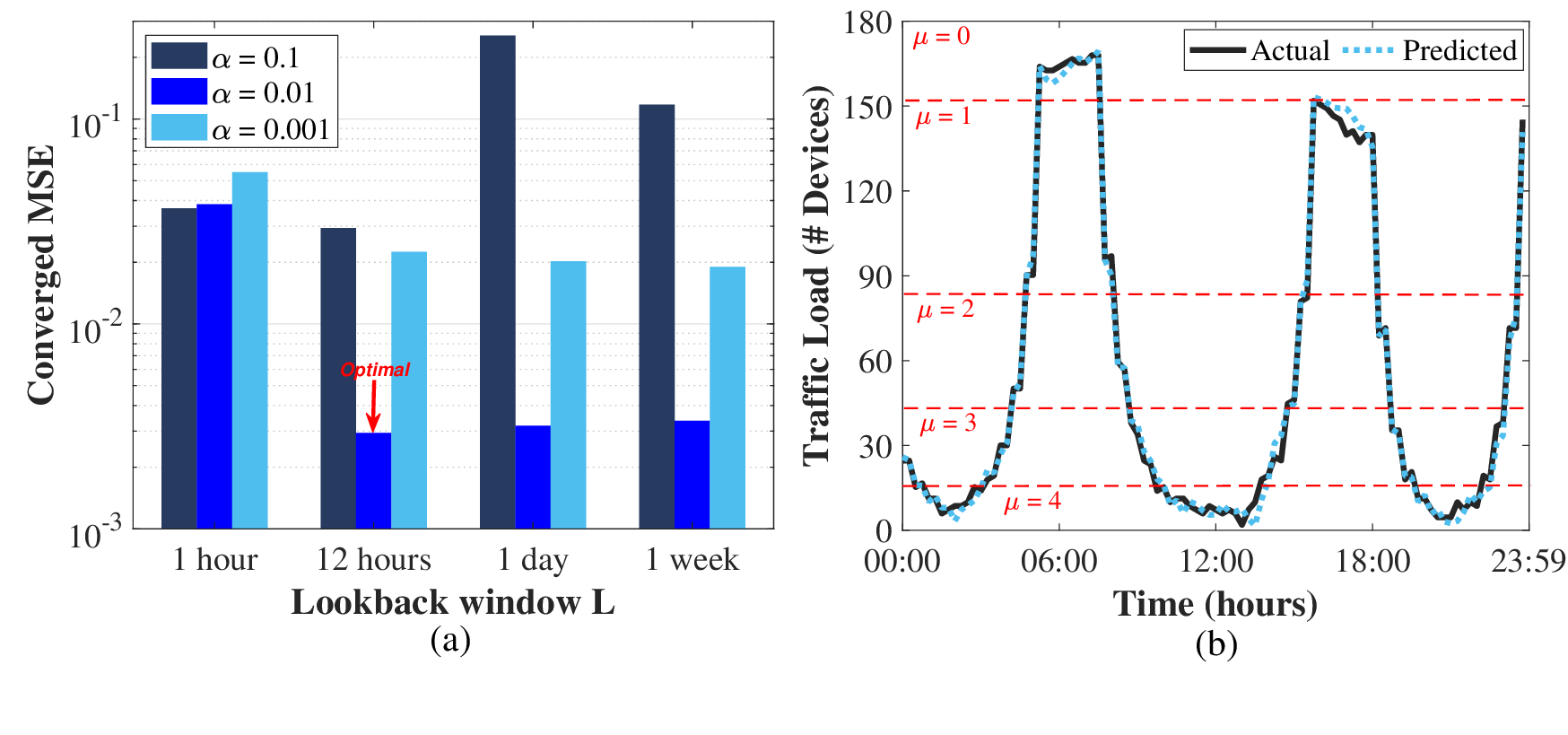}}
\caption{rApp training and inference performance. \textbf{(a)} Traffic prediction final MSE as a function of lookback window $L$ for different learning rates $a$. \textbf{(b)} Actual vs predicted traffic load during the testing (last) day. Decision areas for different selected numerology $\mu$ are separated with the red dashed lines.}
\label{fig:res1}
\end{figure}

This subsection evaluates the performance of the proposed rApp/xApp by separately assessing the rApp training/inference and the xApp allocation efficiency. Following a grid-search of the learning parameters, Fig.~\ref{fig:res1}(a) reports the converged mean squared error (MSE) of the rApp traffic prediction model as a function of the lookback window $L$ for different learning rates $\alpha$. The results show that intermediate temporal contexts (e.g., $L=12$ hours or $48$ samples) combined with the learning rate $a=0.01$ achieve the lowest prediction error ($<0.002$). Fig.~\ref{fig:res1}(b) illustrates the rApp inference performance by comparing the actual and predicted aggregate traffic load during the testing (last) day. The predicted load accurately tracks the traffic variations, enabling the rApp to partition the traffic load space into decision regions corresponding to different numerology indices $\mu$. The region-dependent selection of $\mu$ is used to generate policy $\boldsymbol{\pi}_5$, which is combined with $\boldsymbol{\pi}_1$-$\boldsymbol{\pi}_4$ to configure the xApp every 15 minutes.

\begin{figure}[t]
\centerline{\includegraphics[trim={0.2cm 0.50cm 2.1cm 0.8cm},clip,width=\columnwidth]{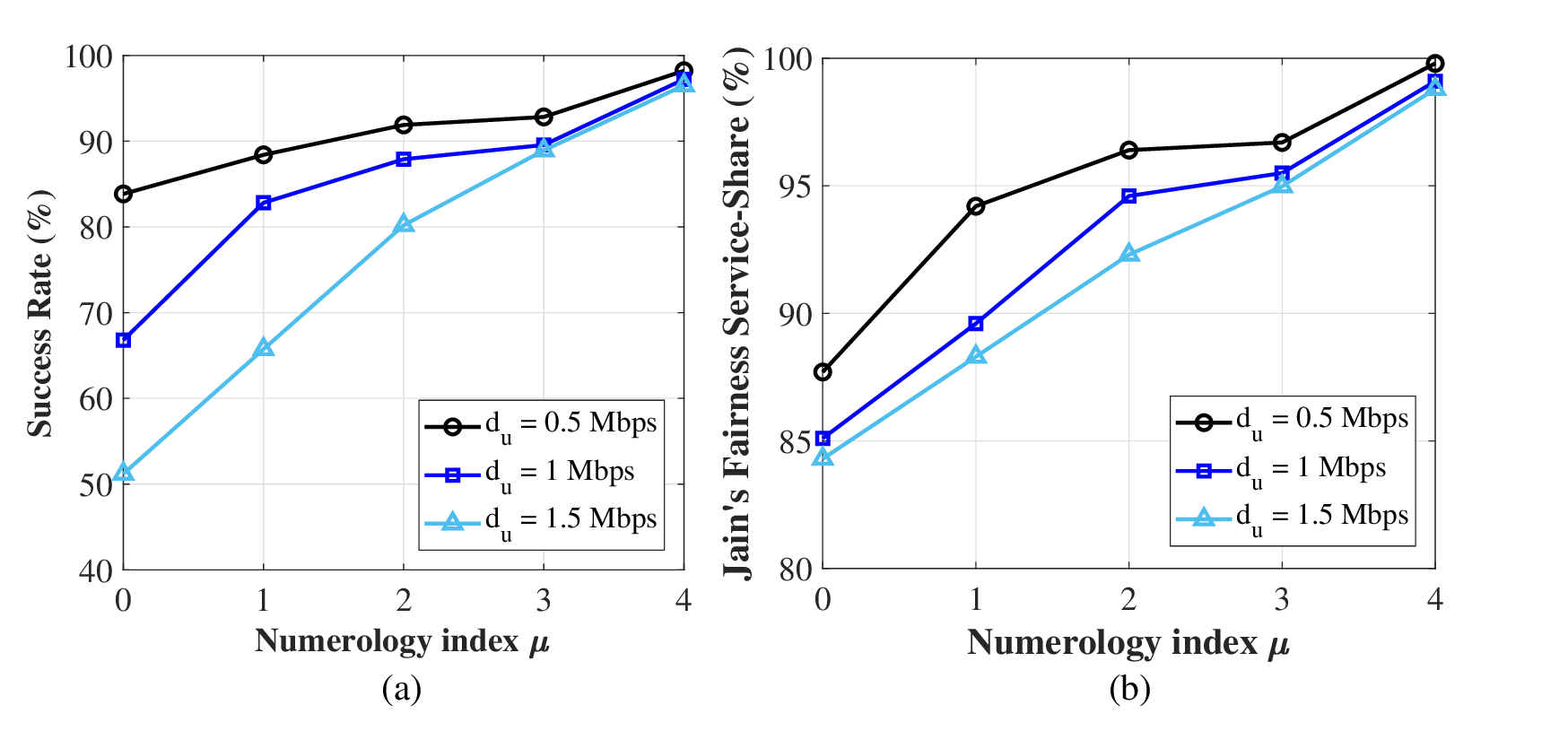}}
\caption{xApp inference performance. \textbf{(a)} Success rate (\%) and \textbf{(b)} Jain's fairness service-share (\%) as a function of the selected numerology $\mu$ for different UE demands $d_u$.}
\label{fig:res2}
\end{figure}

Fig.~\ref{fig:res2} evaluates the xApp performance in terms of \emph{Success Rate} and \emph{Jain's Fairness Service-share} as functions of the selected numerology index $\mu$ under different UE data rate demands $d_u$. The Success Rate is defined as the percentage of UEs that are fully satisfied following the xApp-suggested PRB allocation, averaged over all scheduling intervals of the whole testing day. As shown in Fig.~\ref{fig:res2}(a), the Success Rate increases with $\mu$ due to the larger PRB bandwidth $W$, while higher UE demands lead to a more pronounced performance degradation.

\begin{figure}[t]
%\centerline{\includegraphics[trim={0.2cm 7cm 27.1cm 0.8cm},clip,width=0.95\columnwidth]{Figures/fig_comp.eps}}
\centerline{\includegraphics[trim={0cm 4.7cm 7cm 0.5cm},clip,width=0.97\columnwidth]{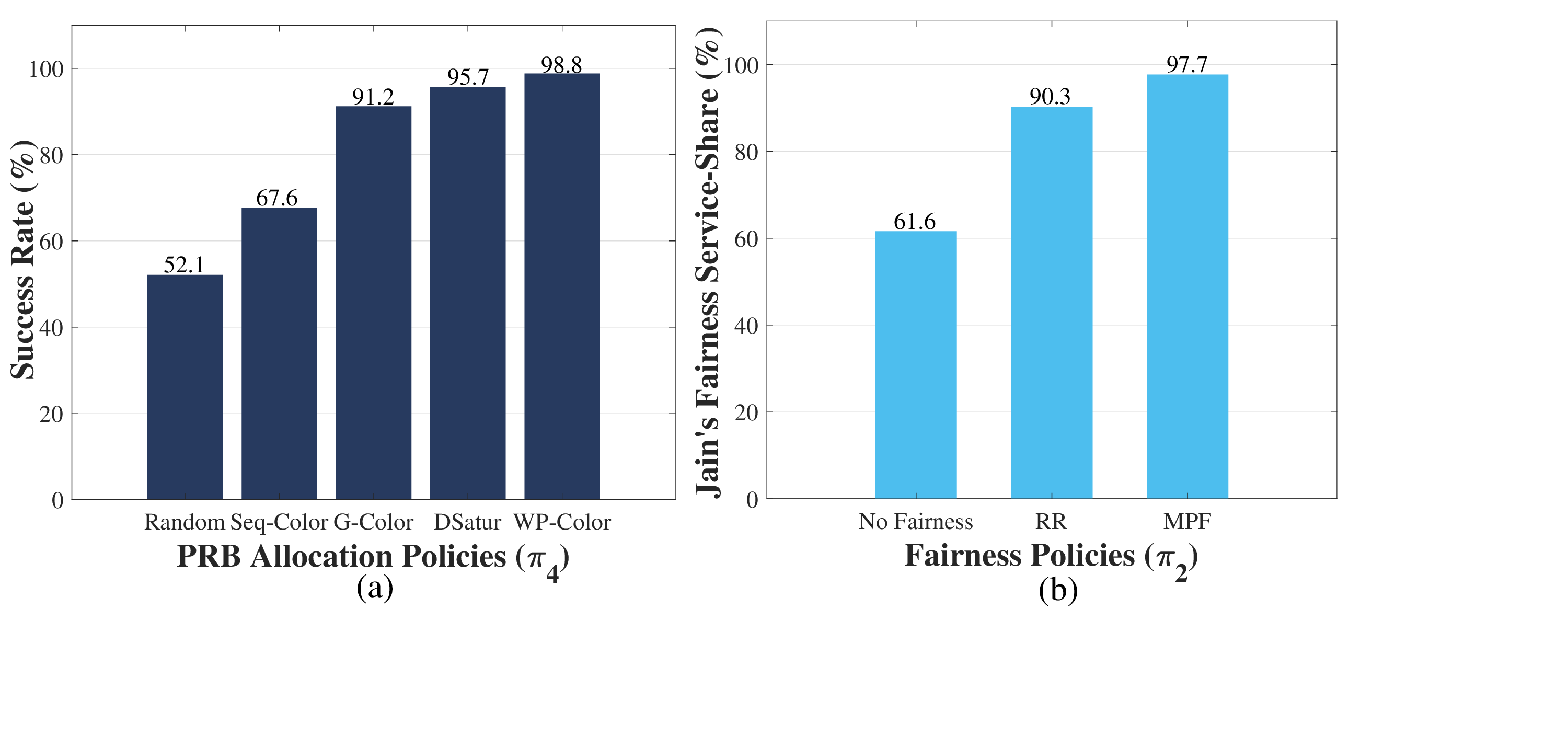}}
\caption{xApp inference performance for different policies. \textbf{(a)} Success rate (\%) for different PRB allocation policies. \textbf{(b)} Jain's fairness service-share (\%) for different Fairness policies.}
\label{fig:res3}
\end{figure}

Service-share fairness is quantified using Jain’s Fairness Index (JFI), computed over the long-term service shares of UEs:

\begin{equation}
\mathrm{JFI} =
\frac{\left(\sum_{u \in \mathcal{U}} \bar{S}_u\right)^2}
{|\mathcal{U}| \sum_{u \in \mathcal{U}} \bar{S}_u^2} \times 100\%,
\end{equation}

\noindent where $\bar{S}_u=\frac{1}{T} \sum_{t=1}^{N_x} \boldsymbol{1}_{\{R_u(t)>d_u\}}$ denotes the fraction of scheduling intervals in which UE $u$ is fully satisfied (function $\boldsymbol{1}_{\{\cdot\}}$ equals 1 when condition $\{\cdot\}$ is true, otherwise 0). Evidently, JFI is a metric in $0-100\%$ that is used to measure PRB allocation fairness among users, with 1 reflecting perfectly fair (equal share) and 0 completely unfair service-share during the $N_x$ xApp runs (time slots). Fig.~\ref{fig:res2}(b) shows that the proposed MPF scheduling significantly improves fairness, presenting at least 85\% JFI even for high number of UEs (i.e., $\mu=0$), while also achieving increasing fairness for higher selected numerology $\mu$ (due to lower number of users). This confirms that the xApp effectively balances interference avoidance, fairness, and PRB utilization.

\subsection{Comparative Analysis}

This subsection evaluates the impact of different xApp policy selections on the performance of the proposed DSA framework. Specifically, we compare alternative policies for the PRB Allocation Engine ($\pi_4$) and the Fairness Engine ($\pi_2$). We consider that all UEs demand $d_u=1$ Mbps and that the system operates at $\mu=4$. 

Fig.~\ref{fig:res3}(a) reports the Success Rate (\%), defined as the fraction of UEs successfully assigned PRBs averaged across all scheduling intervals, for different PRB allocation algorithms. \textit{Random} (each PRB is assigned randomly to a UE) and Sequential Coloring (\textit{Seq-Color}, PRBs are assigned sequentially to UEs, i.e., PRB 1 to UE 1, and so on) schemes exhibit limited performance, achieving Success Rates below $70\%$. Graph-based approaches significantly improve performance, with Greedy Coloring (\textit{G-Color}) \cite{sipayung2022implementation}, \textit{DSatur} \cite{san2012new} and Welsh-Powell Coloring (\textit{WP-Color}) \cite{welsh1967upper} achieving Success Rates above $90\%$, highlighting the importance of interference-aware graph coloring for dense multi-cell deployments.

To assess fairness, Fig.~\ref{fig:res3}(b) also compares JFI for different fairness policies. Without explicit fairness control (Fairness Engine deactivated), service distribution is highly uneven, resulting in a JFI of $61.6\%$. Round-Robin scheduling substantially improves fairness, while the proposed MPF mechanism achieves the highest fairness level ($97.7\%$). This confirms that MPF can effectively balance long-term service equity while remaining compatible with the conflict-aware PRB allocation process. Overall, these results demonstrate that the joint selection of advanced graph coloring ($\pi_4$) and fairness-aware scheduling ($\pi_2$) policies within the xApp is critical to maximizing both service availability and fairness in O-RAN-based spectrum sharing.

\section{Conclusions and Future Extensions}

This paper presented an interoperable rApp/xApp-driven dynamic spectrum allocation framework for O-RAN, leveraging long-term traffic intelligence and near-real-time graph-based optimization. By decoupling traffic prediction and policy generation at the Non-RT RIC from interference-aware PRB allocation and fairness control at the Near-RT RIC, the proposed solution achieves significant gains in PRB assignment success rate and service-share fairness while preserving system stability. The framework is fully aligned with the O-RAN philosophy of functional disaggregation, open interfaces, and multi-timescale control, enabling modular and vendor-agnostic integration of intelligent RAN optimization functions.

Future work will focus on learning-enhanced graph coloring techniques to further improve adaptability under highly dynamic interference conditions, energy-aware extensions that jointly optimize spectrum and power consumption, and cross-domain orchestration across RAN, transport, and cloud resources to support end-to-end optimization in next-generation networks.

\section*{Acknowledgment}

The authors thank the partners from UNITY-6G and 6G-Cloud projects for their collaboration on technical aspects related to this study.

\bibliographystyle{IEEEtran}
\bibliography{mybib}

@article{giannopoulos2025comix,
  title={{COMIX: Generalized Conflict Management in O-RAN xApps-Architecture, Workflow, and a Power Control case}},
  author={Giannopoulos, Anastasios and Spantideas, Sotirios and Levis, George and Kalafatelis, Alexandros and Trakadas, Panagiotis},
  journal={IEEE Access},
  year={2025},
  publisher={IEEE}
}

@inproceedings{akman2024energy,
  title={Energy Saving and Traffic Steering Use Case and Testing by O-RAN RIC xApp/rApp Multi-vendor Interoperability},
  author={Akman, Arda and Oliver, Pablo and Jones, Michael and Tehrani, Peyman and Hoffmann, Marcin and Li, Jia},
  booktitle={2024 IEEE 100th Vehicular Technology Conference (VTC2024-Fall)},
  pages={1--6},
  year={2024},
  organization={IEEE}
}

@inproceedings{giannopoulos2025ai,
  title={{AI-Driven Self-Healing in Cloud-Native 6G Networks Through Dynamic Server Scaling}},
  author={Giannopoulos, Anastasios and Spantideas, Sotirios and Trakadas, Panagiotis and Perez-Valero, Jesus and Garcia-Aviles, Gin{\'e}s and Gomez, Antonio Skarmeta},
  booktitle={2025 IEEE 11th International Conference on Network Softwarization (NetSoft)},
  pages={43--48},
  year={2025},
  organization={IEEE}
}

@article{hoffmann2023open,
  title={Open RAN xApps design and evaluation: Lessons learnt and identified challenges},
  author={Hoffmann, Marcin and Janji, Salim and Samorzewski, Adam and Ku{\l}acz, {\L}ukasz and Adamczyk, Cezary and Dryja{\'n}ski, Marcin and Kryszkiewicz, Pawel and Kliks, Adrian and Bogucka, Hanna},
  journal={IEEE Journal on Selected Areas in Communications},
  volume={42},
  number={2},
  pages={473--486},
  year={2023},
  publisher={IEEE}
}

@article{polese2022colo,
  title={ColO-RAN: Developing machine learning-based xApps for open RAN closed-loop control on programmable experimental platforms},
  author={Polese, Michele and Bonati, Leonardo and D'Oro, Salvatore and Basagni, Stefano and Melodia, Tommaso},
  journal={IEEE Transactions on Mobile Computing},
  volume={22},
  number={10},
  pages={5787--5800},
  year={2022},
  publisher={IEEE}
}

@INPROCEEDINGS{10651558,
  author={Hassan, Manasik and Diab, Ali and Parameswaran, Sriram and Mitschele-Thiel, Andreas},
  booktitle={Mobilkommunikation; 28. ITG-Fachtagung}, 
  title={Machine Learning-Based Coverage and Capacity Optimization xApp/rApp for Open RAN 5G Campus Networks}, 
  year={2024},
  volume={},
  number={},
  pages={191-196},
  doi={}}

@manual{alliance2020,
  title        = "",
  author       = "{{\it O-RAN Architecture Description—V2.00}}",
  organization = "O-RAN Working Group 1",
  address      = "Alfter, Germany",
  year = 2020
}

@article{bonati2021intelligence,
  title={Intelligence and learning in O-RAN for data-driven NextG cellular networks},
  author={Bonati, Leonardo and D'Oro, Salvatore and Polese, Michele and Basagni, Stefano and Melodia, Tommaso},
  journal={IEEE Communications Magazine},
  volume={59},
  number={10},
  pages={21--27},
  year={2021},
  publisher={IEEE}
}

@manual{alliance2020ai,
  title        = "",
  author       = "{{\it AI/ML workflow description and requirements-V01.01}}",
  organization = "ORAN Alliance",
  address      = "Alfter, Germany",
  year = 2020
}

@article{lien20175g,
  title={5G new radio: Waveform, frame structure, multiple access, and initial access},
  author={Lien, Shao-Yu and Shieh, Shin-Lin and Huang, Yenming and Su, Borching and Hsu, Yung-Lin and Wei, Hung-Yu},
  journal={IEEE communications magazine},
  volume={55},
  number={6},
  pages={64--71},
  year={2017},
  publisher={IEEE}
}

@article{chiang20042,
  title={A 2-D random-walk mobility model for location-management studies in wireless networks},
  author={Chiang, Kuo-Hsing and Shenoy, Nirmala},
  journal={IEEE Transactions on vehicular technology},
  volume={53},
  number={2},
  pages={413--424},
  year={2004},
  publisher={IEEE}
}

@article{spantideas2024smart,
  title={{Smart Mission Critical Service Management: Architecture, Deployment Options, and Experimental Results}},
  author={Spantideas, Sotirios T and Giannopoulos, Anastasios E and Trakadas, Panagiotis},
  journal={IEEE Transactions on Network and Service Management},
  year={2024},
  publisher={IEEE}
}

@inproceedings{levis2025sleepy,
  title={{SLEEPY-rApp: Delay-aware sleep scheduling for energy efficiency in MEC-enabled O-RAN}},
  author={Levis, George and Giannopoulos, Anastasios and Spantideas, Sotirios and Trakadas, Panagiotis},
  booktitle={Proceedings of the 2nd International Workshop on MetaOS for the Cloud-Edge-IoT Continuum},
  pages={40--45},
  year={2025}
}

@article{sklar2002rayleigh,
  title={Rayleigh fading channels in mobile digital communication systems. I. Characterization},
  author={Sklar, Bernard},
  journal={IEEE Communications magazine},
  volume={35},
  number={7},
  pages={90--100},
  year={2002},
  publisher={IEEE}
}

@article{zhang2013interference,
  title={Interference graph-based resource-sharing schemes for vehicular networks},
  author={Zhang, Rongqing and Cheng, Xiang and Yao, Qi and Wang, Cheng-Xiang and Yang, Yang and Jiao, Bingli},
  journal={IEEE transactions on vehicular technology},
  volume={62},
  number={8},
  pages={4028--4039},
  year={2013},
  publisher={IEEE}
}

@article{ge20222,
  title={2-layer interference coordination framework based on graph coloring algorithm for a cellular system with distributed MU-MIMO},
  author={Ge, Chang and Xia, Sijie and Chen, Qiang and Adachi, Fumiyuki},
  journal={IEEE Transactions on Vehicular Technology},
  volume={72},
  number={3},
  pages={3557--3568},
  year={2022},
  publisher={IEEE}
}

@article{kim2005proportional,
  title={A proportional fair scheduling for multicarrier transmission systems},
  author={Kim, Hoon and Han, Youngnam},
  journal={IEEE Communications letters},
  volume={9},
  number={3},
  pages={210--212},
  year={2005},
  publisher={IEEE}
}

@article{welsh1967upper,
  title={An upper bound for the chromatic number of a graph and its application to timetabling problems},
  author={Welsh, Dominic JA and Powell, Martin B},
  journal={The Computer Journal},
  volume={10},
  number={1},
  pages={85--86},
  year={1967},
  publisher={Oxford University Press}
}

@article{san2012new,
  title={A new DSATUR-based algorithm for exact vertex coloring},
  author={San Segundo, Pablo},
  journal={Computers \& Operations Research},
  volume={39},
  number={7},
  pages={1724--1733},
  year={2012},
  publisher={Elsevier}
}

@inproceedings{sipayung2022implementation,
  title={Implementation of the greedy algorithm on graph coloring},
  author={Sipayung, TN and Suwilo, S and Gultom, Parapat and others},
  booktitle={Journal of Physics: Conference Series},
  volume={2157},
  number={1},
  pages={012003},
  year={2022},
  organization={IOP Publishing}
}

@article{sevgican2020intelligent,
  title={Intelligent network data analytics function in 5G cellular networks using machine learning},
  author={Sevgican, Salih and Turan, Meri{\c{c}} and G{\"o}karslan, Kerim and Yilmaz, H Birkan and Tugcu, Tuna},
  journal={Journal of Communications and Networks},
  volume={22},
  number={3},
  pages={269--280},
  year={2020},
  publisher={KICS}
}

\end{document}